\documentstyle[aps,preprint,prl,epsfig]{revtex}
\tightenlines

\begin{document}
\draft

\title{Macroscopic chaos in globally coupled maps}

\author{M. Cencini $^{(a)}$, M. Falcioni $^{(b)(\ast)}$, D. Vergni 
        and A. Vulpiani $^{(a)(b)}$}

\address{Dipartimento di Fisica, Universit\`a ``La Sapienza'',
         P.le A. Moro 2, 00185, Roma, Italy}
\address{($\ast$) Corresponding author (fax:+39-6-4463158;
     E-mail: M.Falcioni@roma1.infn.it)}
\address{(a) also INFM, Unit\`a di Roma 1}
\address{(b) also INFN, Sezione di Roma 1}

\date{\today}

\maketitle

\begin{abstract}

We study the coherent dynamics of globally coupled 
maps showing {\it macroscopic chaos}. With this term we indicate the
hydrodynamical-like irregular  behaviour of some global observables, 
with typical times much longer than the times related to the evolution 
of the single (or microscopic) elements of the system. The usual Lyapunov 
exponent is not able to capture the essential features of this macroscopic 
phenomenon. Using the recently introduced notion of finite size Lyapunov 
exponent, we characterize, in a consistent way, these macroscopic behaviours. 
Basically, at small values of the perturbation we recover the usual 
(microscopic) Lyapunov exponent, while at larger values a sort of 
macroscopic Lyapunov exponent emerges, which can be much 
smaller than the former. 
A quantitative characterization of the chaotic motion at 
hydrodynamical level is then possible, even in the absence of the explicit 
equations for the time evolution of the macroscopic observables.

\end{abstract}

\pacs{PACS: 05.45.+b \hfill\break
\noindent KEYWORDS: Macroscopically coherent dynamics, 
Finite size Lyapunov exponent, Globally coupled maps}

\section{Introduction}
In the last years the emergence of non trivial collective 
behaviour in high-dimensional dynamical systems has gathered 
much attention \cite{CM92,Kan95,CFV96,K89,KK91,PK94,Kan96,PC92}.  
An interesting limit case of macroscopic coherence is the global 
synchronization of all the parts of the system: this behaviour 
has been found in many different situations in physics, chemistry 
and biology. Relevant examples of synchronization are the 
flickering of swarms of fireflies, fish locomotion, Josephson 
arrays and certain multi-mode laser systems \cite{W80}. 

Beyond synchronization there exist much more interesting 
and intriguing phenomena involving non trivial coherence.  
Let us mention the relevant works of Kaneko \cite{Kan95} 
showing the emergence of subtle coherence among different 
elements of globally coupled maps (GCM), in the case of a 
logistic local map. 

A GCM system (of the type considered in this paper) is a collection of
elements $x_n$ (defined on the site $n$) evolving according to the
equation:
\begin{equation}
\label{eq:uno}
x_n(t+1)=(1-\epsilon)f_a(x_n(t))+ 
  \frac {\epsilon} {N} \sum_{i=1}^{N}  f_a(x_{i}(t)),
\end{equation}
where $N$ is the total number of elements, $t$ is the discrete
time and $f_a(x)$ is a nonlinear function depending on a 
parameter $a$. 

When the mapping function is $f_a(x)= ax(1-x)$ 
(i.e. a logistic map), for certain values of $\epsilon$ and $a$ 
one can observe rather remarkable behaviours, as anomalous fluctuations, 
glassy behaviour and broken ergodicity \cite{CFV96}. 
For instance, Kaneko \cite{Kan95} studied a parameters region
where the number of positive Lyapunov exponents is proportional to $N$; 
in this region the local elements $x_n$ take different values, 
almost randomly, so one might expect that the temporal fluctuations 
of the mean-field
\begin{equation}
\label{eq:due}
h(t)= \frac {1}{N} \sum_{i=1}^{N}  f_a(x_{i}(t))
\end{equation}
obey a Gaussian statistics, with a mean square deviation 
$\sigma _{N}^2 =\langle h^2 \rangle - \langle h \rangle^2$ 
proportional to $N^{-1}$. On the contrary one observes 
$\sigma _{N}^2 \to {\rm const}$, as $N \to \infty$. Only with 
the inclusion of noise one obtains $\sigma _{N}^2 \sim N^{-\beta}$, 
with $\beta < 1$, if the noise is not too strong. 
This is due to a coherence (collective behaviour) among the 
elements $x_n$, that can be associated \cite{Kan95} with the 
existence of the periodic windows in the chaotic region of the 
logistic map.

\noindent In some regions of the parameters space one observes 
clustering \cite{K89}, i.e., the elements $x_n$ split into clusters, 
such that the elements belonging to the same cluster $k$ take an 
identical value $\tilde{x}_k (t)$. The number ${\cal C}$ of the possible 
clusters can be very large with $N$ ($\ln {\cal C} \sim \sqrt{N}$) 
and the dynamics can show glass-like behaviour and failure of 
ergodicity \cite{CFV96}. 

Symplectic GCM, where the coupling is not of the mean-field type 
of eq. (\ref{eq:uno}), 
can show a more intriguing behaviour for the clustered motion, since, 
in this case, the synchronization between the different parts is not 
perfect and the clustering has a finite lifetime \cite{KK91}.

Even coupled map lattices (CML) with local coupling, on 
$d$-dimensional lattices, can show a macroscopic behaviour that 
is very different from the microscopic one. For example, Chat\'e 
and Manneville \cite{CM92} showed that in CML on hyper-cubic 
$d$-dimensional lattices some macroscopic variables, as the center 
of mass:
\begin{equation}
\label{eq:tre}
m(t)= \frac {1} {N} \sum_{i=1}^{N} x_{i}(t),
\end{equation}
can have periodic behaviour, or quasi-periodic for $d\geq 5$, 
even if the system is chaotic, i.e. the first Lyapunov exponent 
is positive. 

In our opinion, one of the more interesting facts among the 
many different aspects of the high-dimensional dynamical systems 
is the {\it macroscopic chaos}. With this locution we indicate the 
irregular behaviour of some global variables (e.g. the center 
of mass) that evolve with typical times much longer than the 
characteristic time of the full dynamics, that we call the 
microscopic dynamics; the order of magnitude of the latter 
time may be estimated by the inverse of the first Lyapunov 
exponent $\lambda_1$ of the microscopic dynamics 

The macroscopic chaos for the  high-dimensional maps, in some sense, 
is the analogous of the hydrodynamical chaos for the molecular 
motion. Let us briefly clarify this point and stress the open problems.
For the Hamiltonian dynamical system describing the evolution of the 
molecules of a fluid the order of magnitude of the microscopic 
characteristic time is the mean collision time. On the other hand, 
if one studies the hydrodynamic equations of the considered fluid 
-- those describing the motion on space and time scales that are very 
large with respect to the molecular scales -- the involved 
characteristic times are much larger. So one expects that for 
the full Hamiltonian dynamics $\lambda_1 \sim 10^{11}\quad {\rm s} ^{-1}$, 
while the first Lyapunov exponent for the hydrodynamic equations, 
in a chaotic regime, is of order $(10^{-1} - 1)\quad {\rm s} ^{-1}$, 
the exact value depending on the geometry of the system, the external 
forcing, the Reynolds number and so on. 

This big difference comes from the coarse graining process 
in the building of the hydrodynamic equations, that is from 
the very different scales one is observing in the two cases.

If one is able to write down the equations at the hydrodynamical level 
(i.e. for the slow variables) then it is not difficult to characterize 
the macroscopic behaviour, by means of the standard techniques of 
dynamical system theory. However in this case, one subtle point, on 
the mathematical side, is the problem of the norm to be used in the 
definition of the Lyapunov exponents. For finite dimensional systems 
all the norms are equivalent \cite{Kol} and there is no ambiguity. 
This is not true for infinite dimensional systems (as the hydrodynamical 
partial differential equations are), so that the Lyapunov exponents 
can depend on the used norm. 

In generic CML, as far as we know, there are no general 
systematic methods to build up the macroscopic equations; 
therefore it is necessary to study the macroscopic behaviour 
of the system in terms of the $x_n$ variables, i.e. it is necessary 
to remain at the microscopic level of description. 

The present paper is organized as follows. In the next section we
report some phenomenology of macroscopic chaos in GCM. Section \ref{sec:3} is
devoted to a discussion on the recently proposed self-consistent
Perron-Frobenius method for the one-body description of these systems
at a macroscopic level. In section \ref{sec:4}, that is the nucleus of 
the paper, we show that the recently introduced {\it finite size 
Lyapunov exponent} gives a suitable
quantitative characterization of the macroscopic chaos; that is to
say: we are able to identify a macroscopic characteristic time by
means of a Lyapunov exponent that is effective on the macroscopic
scales and which, in general, is much smaller than the usually defined
one.  The last section is devoted to discussion and conclusions.  In
the Appendix we discuss some subtle points related to the problem of
the choice of the norm in the definition of the Lyapunov exponent for
the one-body description.
       
\section{Some phenomenological results and first naive 
  approaches}
\label{sec:2}

For sake of self-consistency, we report some numerical evidence 
(partly already known from previous works \cite{Kan95,PK94,Kan96})
of the existence of macroscopic chaos in GCM, eq. (\ref{eq:uno}). 
For a detailed description of the many different macroscopic 
behaviours see the works of Kaneko \cite{Kan95,Kan96} and also 
Pikovsky and Kurths \cite{PK94}.

We studied mainly systems (\ref{eq:uno}) of globally coupled tent 
maps, i.e.: $f_a (x) = a(1/2 - |x-1/2|)$, and all the figures of this 
paper refer to these systems. In addition, to test the validity 
of our method, we also considered a slightly different version of 
eq. (\ref{eq:uno}), the so-called heterogeneous GCM \cite{Kan96}, 
in which the interacting elements are not all identical:
\begin{equation}
\label{eq:dueuno}
x_n(t+1)=(1-\epsilon)f_{a_n}(x_n(t))+ 
  \frac {\epsilon} {N} \sum_{i=1}^{N}  f_{a_i}(x_{i}(t)),
\end{equation}
the control parameters $a_i$ are quenched variables that 
take values according to some rule to be specified. Here we chose 
$f_{a_i} (x)= 1- a_{i} x^2$ (the logistic map), with $a_i$ assuming 
equally spaced values in an interval between $a_{min}$ and $a_{max}$. 

At varying $\epsilon$ and $a$ in eq. (\ref{eq:uno}) or $a_{min}$ and 
$a_{max}$, in eq. (\ref{eq:dueuno}), and by looking at the evolution 
of a macroscopic variable as, for instance, $m(t)$, we observed 
different macroscopic behaviours, which can be grouped in the three 
following classes.
\begin{description} 
\item[i)] STANDARD CHAOS: the microscopic Lyapunov 
exponent is positive and the law of large numbers holds. In 
this case, if one plots $m(t)$ versus $m(t-1)$, one observes a 
spot with size of order $1/\sqrt{N}$.
\item[ii)] MACROSCOPIC PERIODICITY: the microscopic Lyapunov 
exponent is positive and the plot of $m(t)$ versus $m(t-1)$ gives 
rise to $k$ spots with size of order $1/\sqrt{N}$, for a macroscopic 
motion of period $k$. There are also less trivial periodic or 
quasi-periodic behaviours (e.g. in \cite{Kan96} a case is shown in 
which the mean-field dynamics evolves on a 2-dimensional torus).
\item[iii)] MACROSCOPIC CHAOS: the microscopic Lyapunov exponent is 
positive and the plot of $m(t)$ versus $m(t-1)$ shows an apparently 
well structurated function, see fig. 1, that suggests chaotic 
motion for $m(t)$.
\end{description}
The first hope is that, at a macroscopic level, the system can be
described by few global variables, that are able to catch the large
scale properties, i.e. one hopes that the macroscopic motion is driven
by a low dimensional dynamics. Indeed, fig. 1a (that is obtained for
particular parameters values of the tent GCM), seems to suggest such a
possibility: the first-return plot for the center of mass evolution
appears to generate a one dimensional map. In such a case one can
conjecture that
\begin{equation}
\label{eq:duedue}
m(t+1) = F(m(t)) + \eta_N (t),
\end{equation}
where $\eta_N$ is a small ($\sim 1/\sqrt{N}$) incoherent 
fluctuation, vanishing in the limit $N\to\infty$. If the above 
equation holds -- with the evolution function $F(m)$ 
fitted from the data -- one can define the
Lyapunov exponent for dynamical system (\ref{eq:duedue}). This 
latter quantity gives information on the sensibility of the 
evolution of the macroscopic variable $m$ on the initial 
conditions, so that one may call it {\it macroscopic} Lyapunov 
exponent $\lambda_{macro}$; according to the standard definition, 
one has: 
\begin{equation}
\label{eq:duetre}
\lambda_{macro} = \int \ln |F'(m)| P(m) dm,
\end{equation}
where $P(m)$ is the probability distribution of $m$, as obtained 
from numerical simulations of eq. (\ref{eq:uno}). 
In the case of macroscopic chaos one expects that 
$\lambda_{macro} > 0$. On the contrary, using the recipe of eq. 
(\ref{eq:duetre}) one obtains that $\lambda_{macro}$ is negative. 
This result is not a paradox: simply eq. (\ref{eq:duedue}) does 
not hold. If one looks carefully at the plot of $m(t)$ versus $m(t-1)$, 
for large $N$, one realizes that very fine grained structures 
exist, see fig. 1b. This suggests that the variable $m(t)$ alone is 
not sufficient for a satisfying description of the macroscopic 
properties of the system. Therefore we need to introduce some other 
variables. It seems likely that the system evolution at a macroscopic 
scale is driven by a full set of equations of phenomenological type, 
such as, e.g.:
\begin{equation}
\left \{ \begin{array}{lll}
 m(t+1) & = & F_1 (m(t), \sigma (t), \dots )+\eta^{(1)}_{N} 
 \\ 
    \sigma (t+1) & = & F_2 (m(t), \sigma (t), \dots )+\eta^{(2)}_{N}  \\
    \vdots &   & \vdots  
\end{array} \right.
\label{eq:duequattro}
\end{equation}
where $\sigma ^2 = (\sum x_i ^2)/N -(\sum x_i/N)^2$, and $\eta^{(i)}_{N}$ are
small fluctuations vanishing as $N \rightarrow \infty$. 
Unfortunately, as far as we know, there is not a systematic approach 
(even of numerical type) for the building of eq.s (\ref{eq:duequattro}),
i.e. for the determination of the functions $F_1, F_2, \dots$. 

Another naive approach, based on the possible {\it time scales 
separation} between the microscopic and the macroscopic dynamics, 
is the analysis of the microscopic and macroscopic time 
correlation functions, defined as follows:  
\begin{equation}
\label{eq:duecinque}
C^{(i)} _{micro} (\tau) ={\langle x_i (t+\tau) x_i (t) \rangle - 
\langle x_i \rangle ^2  \over 
\langle x^2 _i \rangle - \langle x_i \rangle ^2} 
\end{equation}
\begin{equation}
\label{eq:duesei}
C _{macro} (\tau) = {\langle m (t+\tau) m (t) \rangle - 
\langle m \rangle ^2  \over 
\langle m^2  \rangle - \langle m \rangle ^2}, 
\end{equation}
where $\langle(...)\rangle$ means time average along the trajectory.

It is to be noted, at this point, that the system we are studying 
shows a non trivial macroscopic behaviour in a parameter region where 
it is also in a broken ergodicity phase. In fig. 2 we show the 
probability distributions, $P_{n}(x)$, of two different $x_{n}$ for 
the system (\ref{eq:uno}) in a situation of macroscopic chaos. It is 
well evident that the knowledge of the probability distribution of 
a generic $x_{n}$ does not permit a good description of the system 
(this is what we mean, in such a setting, by saying that the system is 
not ergodic). It is important to note that the two elements have been 
chosen in the two different bands in which the $\{x_{n}\}$ are grouped, 
see \cite{Kan95}. Moreover elements belonging to the same band seems 
to have the same distributions. Let us stress that this behaviour is 
not related to the particular band structure of the tent GCM: we found 
non trivial macroscopic motion in the presence of broken ergodicity also 
in heterogeneous GCM (\ref{eq:dueuno}), where the band structure is 
absent, but the microscopic building elements are not identical. 

Fig. 3 shows the above defined correlation functions 
in the regime of macroscopic chaos. Since one has a failure 
of the ergodicity, different sites have different autocorrelations,
that is $C^{(i)} _{micro} (\tau)$ depends on the band to which the
element $x_{i}$ belongs. Rather unexpected is the fact that the 
characteristic decay times of $C^{(i)} _{micro} (\tau)$, 
$\tau_m ^i$, and of $C _{macro} (\tau)$, $\tau_M$,
defined by the equations 
\begin{equation}
C^{i} _{micro} (\tau) \sim e^{-\tau/\tau_m ^i}, \qquad 
C _{macro} (\tau) \sim e^{-\tau/\tau_M}, 
\end{equation}
are not very different; for instance, in the case of fig. 3 one has: 
$\tau_m ^i \approx \tau_M \approx 115$. A non trivial 
point is that these typical decay times are much larger 
than the characteristic time given by the 
Lyapunov exponent, i.e. $1/\lambda_{1}$ (for the system of fig. 3 
one has $1/\lambda_{1} \approx 4.$). This fact can be related to 
the strong intermittency of the system. We have to stress that 
this behaviour of the correlation times has been observed also in the 
heterogeneous logistic GCM (\ref{eq:dueuno}).

In the next sections we discuss some less naive approaches to 
the characterization of the macroscopic chaos.

\section{One-body description in terms of a suitable 
         Perron-Frobenius operator}
\label{sec:3}

Let us now discuss how one can try to characterize
the coherent dynamics of GCM in terms of a suitable 
Perron-Frobenius (PF) operator \cite{Kan95,PK94,PC92}. 
At any time $t$ one can wonder about the snapshot density 
$\rho _t (y)$, defined by 
the probability density that $x_n (t)$ assumes the value $y$:
\begin{equation}
\label{eq:treuno}
\rho_t(y)= \frac {1} {N} \sum_{i=1}^{N} \delta (y-x_{i}(t)).
\end{equation}
Using the fact that, for a mean field coupling as in eq. 
(\ref{eq:uno}), each $x_n$ interacts with all the other 
$\lbrace x_i\rbrace $ in the same way, the time evolution of 
$\rho_t(y)$ can be written in terms of $\rho_{t-1}(y)$, by 
means of a self-consistent approach:
\begin{equation}
\label{eq:tredue}
\rho_t(y)= {\cal L}^{(1)} \rho_{t-1}(y) = 
\frac {1} {1-\epsilon} \sum_{i} \frac {\rho_{t-1}(y_i)} 
{|f'_a (y_i)|},
\end{equation}
where the sum is on the preimages given by 
\begin{equation}
\label{eq:tretre}
y_i = f^{-1} _a \Biggl( \frac 
          {y- \epsilon\int f_a(z) \rho_{t-1}(z) dz } 
           {1-\epsilon}\Biggr), 
\end{equation}
$f^{-1} _a$ being the inverse function of $f_a$,
for instance, if $f_a (y)= 1-ay^2$ then 
$f^{-1}_a (z)= \pm \sqrt{(1-z)/a}$.

In the parameter region of the standard chaos (see sect. \ref{sec:2}) 
where $m(t)={\rm const.} + O(1/\sqrt{N})$, one has that, after 
long times, $\rho_{t}(y)$ converges toward a stationary 
distribution $\rho_{eq}(y)$, i.e.:   
\begin{equation}
\rho_{t}(y) \to  \rho_{eq} (y) \qquad 
{\rm for} \quad t \to \infty .  
\end{equation}
On the other hand, for a macroscopic periodic behaviour, 
the long time properties of the density $\rho_t (y)$
are described by means of a finite collection of functions 
of $y$; e.g. when $m(t)$ asymptotically oscillates among two 
different values: 
\begin{equation}
m(2t)= m_e  \qquad m(2t+1)=m_o, 
\end{equation}
after a transient time one observes 
\begin{equation}
\rho_{2t} (y) = \rho_e (y) \qquad \rho_{2t+1} (y)=\rho_o (y). 
\end{equation}
Finally, in the case of macroscopic chaos $\rho_{t} (y)$ does 
not converge toward any finite set of function of $y$, but 
evolves following a nontrivial time behaviour. 

The above approach for $\rho_{t} (y)$ has been introduced by 
Perez and Cerdeira \cite{PC92} and K. Kaneko \cite{Kan95}. From 
a conceptual point of view eq.s (\ref{eq:tredue} -- \ref{eq:tretre}) 
are close to the Boltzmann equation for the gases \cite{Kreuz81}. 
Indeed, as one can check by a direct inspection, the PF operator 
${\cal L}^{(N)}$ that rules the evolution of the probability density 
$\rho_t ^{(N)} (y_1, y_2,\dots,y_N)$ in the whole phase space:
\begin{equation}
\rho_t ^{(N)} (x_1, x_2,\dots,x_N) = {\cal L}^{(N)} 
            \rho_{t-1} ^{(N)} (x_1, x_2,\dots,x_N), 
\end{equation}
is a linear operator. On the other hand the PF operator 
${\cal L}^{(1)}$ for the evolution of $\rho_{t} (y)$, 
given by eq.s (\ref{eq:tredue} -- \ref{eq:tretre}), 
is a nonlinear one \cite{PK94}, since it depends explicitly on 
$\rho_{t-1} (y)$. In the kinetic theory of gases one has 
a similar situation: the Liouville equation, for the evolution 
of the probability density in the whole phase space, is 
linear, while the Boltzmann equation for the one-body 
statistics is nonlinear. 

Since the evolution equations (\ref{eq:tredue} -- \ref{eq:tretre}) 
for $\rho_{t} (y)$ are a dynamical system (though of infinite dimension), 
one can formally introduce the notion of first Lyapunov 
exponent \cite{Kan95}. By so doing, as noted in the introduction, 
one is faced with the mathematical problem of the choice of the norm: 
for sake of simplicity we adopt the usual norm in the $L_2$ space.
One could think that the problem of the norm is relevant just for 
strictly mathematical reasons, while from a physical point of view 
this problem practically does not exist. In the Appendix we  
discuss how even for very simple shapes of $\rho(x)$ one obtains
different Lyapunov exponents using different norms. 
 
However to practically obtain the first Lyapunov exponent, 
firstly one applies an infinitesimal perturbation $\delta \rho_0 (y)$ 
on $\rho_{t} (y)$ (in order to satisfy the normalization condition 
one has to impose the constraint $\int\delta \rho (y) dy=0$) then one 
observes the evolution $\delta \rho_n (y)$ of the perturbation. 
In this way, following the standard methods of Benettin et al. 
\cite{BGGS80}, it is possible to estimate the whole 
set of the Lyapunov exponents 
$\lbrace\tilde{\lambda}_i\rbrace$; in particular one has 
\begin{equation}
\label{eq:trequattro}
\tilde{\lambda}_1 = {1 \over 2}\lim _{T\to \infty} 
 \lim _{\delta \rho_0 \to 0} \frac {1} {T}
 \sum_{n=0}^{T-1}  \ln 
 \frac {\int |\delta \rho_{n+1} (y)|^2 dy} 
      {\int |\delta \rho_{n} (y)|^2 dy}.
\end{equation}

It is easy to understand that $\tilde{\lambda}_1$ can give 
interesting ``global information'' which is completely different 
from the one given by the microscopic Lyapunov exponent. In the 
case of standard chaos, or macroscopic periodic behaviour, 
$\tilde{\lambda}_1$ is negative, or zero, while for macroscopic 
chaos $\tilde{\lambda}_1 > 0$. Let us stress that $\tilde{\lambda}_1$ 
can be negative even if the first Lyapunov exponent for the dynamical 
equations (\ref{eq:uno}) is positive. Some detailed numerical 
investigation of Kaneko on GCM show that $\tilde{\lambda}_1$ is 
positive in the cases of macroscopic chaos \cite{Kan95}. 

Equations (\ref{eq:tredue} -- \ref{eq:tretre}) 
hold only in the case of a mean field interaction for homogeneous 
GCM (eq. \ref{eq:uno}) and other particular systems \cite{PK94}, 
but they are not valid even for the 
heterogeneous GCM (eq. \ref{eq:dueuno}). Nevertheless, the 
computation of $\tilde{\lambda}_1$ according to the definition 
(\ref{eq:trequattro}), remains possible using a numerical 
approach -- at least in principle, even if it may result in a 
very huge and heavy task -- also for generic CML (i.e., in the 
case of maps with local coupling). This is accomplished by 
following the evolution of $\rho_{t} (y)$, as given by the time 
development of the histogram of the instantaneous microscopic 
configuration $(x_1, x_2,\dots,x_N)$. 

Besides the previous relevant properties of $\tilde{\lambda}_1$, 
we have to underline a delicate point that rises for its numerical 
computation, when it is not possible to write down the appropriate 
eq. (\ref{eq:tredue}). In this case, as we already pointed, one is 
forced to study the evolution of $||\delta \rho_{t} (y)||$ making 
histograms from two instantaneous configurations 
$\lbrace x_i \rbrace$ and $\lbrace x^{'}_{i} \rbrace$, 
obtained by the introduction of a small perturbation on 
$\lbrace x_i \rbrace$ at $t=0$. To show where is the problem, 
let us consider the time evolution of $\delta m(t)$ in the case of 
standard chaos. One can write $\delta m$ in two different ways:
\begin{equation}
\label{eq:trecinquea}
\delta m(t) = \frac {1} {N}  \sum_{n=1}^{N} \delta x_n (t)
\end{equation}
\begin{equation}
\label{eq:trecinqueb}
\delta m(t) =  \int y \delta \rho_t (y) dy .
\end{equation}
If we are in the parameters region where the Lyapunov 
exponent $\lambda _1$ is positive, but 
$\tilde{\lambda}_1$ is negative, we have both 
$|\delta x_n (t)| \sim |\delta x_n (0)|\exp (\lambda _1 t)$ and 
$||\delta \rho_t|| \sim ||\delta \rho_0|| \exp (-|\tilde{\lambda}_1|t)$. 
So that apparently we fall into a contradiction, since from 
(\ref{eq:trecinquea}) we obtain   
\begin{equation}
\label{eq:treseia}
|\delta m(t)| \sim e ^{\lambda _1 t} ,
\end{equation}
while from (\ref{eq:trecinqueb}) we have  
\begin{equation}
\label{eq:treseib}
|\delta m(t)| \sim e ^{-|\tilde{\lambda}_1|t } .
\end{equation}
To eliminate the contradiction one needs to take into 
proper account the scales of the perturbations on which these two 
equations hold true; eq. (\ref{eq:treseia}) holds at 
small scales: $|\delta m| \ll O(1/\sqrt{N})$, while eq. 
(\ref{eq:treseib}) holds for $|\delta m| \gg O(1/\sqrt{N})$.
Basically if one starts with a perturbation $\delta m(0)$ 
much smaller than $O(1/\sqrt{N})$ one has an exponential growth, 
but only up to a magnitude $O(1/\sqrt{N})$, the maximum allowed by 
the laws of large numbers. In an analogous manner starting from 
a perturbation $\delta m(0)$ much larger than $O(1/\sqrt{N})$ one 
has an exponential decrease up to $O(1/\sqrt{N})$, see fig. 4.

 From the previous example it is clear that if one studies 
numerically the time evolution of $||\delta \rho _t||$ for large 
but finite $N$, at small times (or, more precisely, small 
$||\delta \rho||$) one can observe just the ``microscopic'' 
Lyapunov exponent. So one has to introduce some coarse-graining 
-- i.e. to consider non infinitesimal $\delta \rho _0 $ -- to 
obtain a macroscopic level description. However, it must be noted
that there is a large arbitrariness in choosing a non infinitesimal 
$\delta \rho _0 $. 

In this section we have discussed the computation of 
the macroscopic Lyapunov exponent by means of the 
Perron-Frobenius method, both in its original formulation 
eqs. (\ref{eq:tredue}-- \ref{eq:tretre}) and in its numerical version, to 
be applied when the P-F equation cannot be explicitly written. 
Neither approach is easy to handle because of the arbitrariness 
in the choice of the norm and of the non infinitesimal $\delta \rho _0 $.

In order to avoid these difficulties (numerical and conceptual)
concerning the one-body description, in the next section 
we introduce a different approach.

\section{Characterization of the macroscopic chaos
         {\it via} the predictability in the large}
\label{sec:4}

The introduction of efficient methods (Benettin et al. \cite{BGGS80})
for the computation of the Lyapunov exponents has been an important
milestone in the understanding of chaotic systems.  However, in spite
of their deep conceptual relevance, the Lyapunov exponents are not
able to give a complete characterization of realistic dynamical
systems, such as systems with intermittent behaviour or with many
different characteristic times. For example, in order to describe
intermittency effects one needs to introduce generalized
Lyapunov exponents and the statistics of the fluctuations of the
effective Lyapunov exponent \cite{PV87,EP86,KB91}. 
On the other hand in systems with
many different characteristic times (e.g. the fully developed
turbulence) the Lyapunov exponent cannot be used to study, even at
very rough level, the predictability problem for small (but not
infinitesimal) perturbations, that cannot be described in terms of the
tangent vector. In order to overcome these difficulties recently a new
indicator has been introduced: the finite size Lyapunov exponent
(FSLE), see Ref. \cite{ABCPV96} for details.

The definition of FSLE may be given in terms of the predictability 
time $T_{r}(\delta)$, that is the time a perturbation of initial 
size $\delta$ takes to grow by a factor $r$ ($>1$) during the
system evolution. The perturbation of size $\delta$ is supposed to be 
already aligned with the most unstable direction. The factor $r$ 
should be taken not too large, in order to avoid the growth trough 
different scales. In many applications, $r=2$, so that sometimes the 
$T_{r}(\delta)$ is also called the error doubling time. The Finite 
Size Lyapunov Exponent is defined from a suitable average of the 
predictability time according to 
\begin{equation}
\lambda(\delta)=\left< 
		      \frac{1}{T_{r}(\delta)}\right >_{{\cal N}} \ln \,r =
		      \frac{1}{\langle T_{r}(\delta)\rangle} \ln \,r
		\label{fsle}
\end{equation}
where $\langle\cdots\rangle _{{\cal N}}$ denotes average with the natural 
measure along the trajectory and $\langle\cdots\rangle$ is the 
average over many realizations, for more details see ref. \cite{ABCPV97}.
The above definition holds for continuous time system. 
In the case of discrete time systems (i.e. maps)
$T_{r}(\delta)$ is the minimum time such that the distance $\Delta_{r}$
between the two realizations is greater or equal to $r \delta$.
Therefore, instead of eq. (\ref{fsle}) we have \cite{ABCPV97}:
\begin{equation}
\lambda(\delta)=\frac{1}{\langle T_{r}(\delta)\rangle } 
\left< \ln \left( \frac{\Delta_{r}}{\delta}\right)\right >\,. 
\label{fsle:map}
\end{equation}

Let us note that by the definition (\ref{fsle}) $\lambda(\delta)$
is not appropriate in discerning cases with $\lambda=0$ and $\lambda<0$,
due to the positiveness of the predictability time.

For chaotic systems the limit of infinitesimal perturbations 
$\delta \rightarrow 0$ in (\ref{fsle}) gives the maximum 
Lyapunov exponent $\lambda_{1}$, i.e.  $\lambda(\delta)$
displays a plateau at the value $\lambda_{1}$ for sufficiently
small $\delta$.

In many realistic situations, the error growth for infinitesimal 
perturbations is dominated by the faster scales, that are 
typically the smaller ones (as in the classic example of three 
dimensional turbulent flows). 
When the size $\delta$ of the perturbation cannot be considered any 
longer infinitesimal, the behaviour of 
$\lambda(\delta)$ is governed by the nonlinear evolution of the 
perturbation, and, in general, $\lambda(\delta)<\lambda_{1}$.
The decrease of $\lambda(\delta)$ does follow a system-dependent
law. In some cases, $\lambda(\delta)$ can be predicted by dimensional 
arguments, e.g. in the fully developed turbulence one has
the universal law $\lambda(\delta) \sim \delta^{-2}$, in the inertial range 
\cite{ABCPV96,ABCPV97}.

Therefore, the behaviour of $\lambda$ as a function of $\delta$ gives 
important information on the characteristic times (and scales 
\cite{ABCCV97}) governing the system and is a powerful tool 
in studying dynamical systems which involve many characteristic 
scales in space and time.

The FSLE has been successfully used even in time series analysis 
\cite{BCPPV98}.
A relevant practical fact is that, at variance with the usual Lyapunov 
exponent which needs high embedding dimensions and extremely huge statistic,
the FSLE at relatively large $\delta$ can be extracted even using moderately 
small embedding dimension. For example, in systems in which some slow variables
 can be separated from the fast variables, the FSLE related to the slow 
dynamics can be obtained even with moderate statistics and unresolved small 
scales (i.e. the fast ones).

To practically compute the FSLE, firstly one has to define a series
of thresholds $\delta_{n}=r^{n}\delta_{0}$ ($n=1,\dots, M$), 
and to measure the times $T_{r}(\delta_{n})$ that a perturbation of 
size $\delta_{n}$ takes to grow up to $\delta_{n+1}$. The times 
$T_{r}(\delta_{n})$ are obtained by following the evolution of a 
perturbation from its initial size $\delta_{min} < \delta_0$ up to 
the largest threshold $\delta_{M}$. This can be done by 
integrating two trajectories of the system starting at an initial 
distance $\delta_{min}$. The FSLE, $\lambda(\delta_{n})$, is then 
computed by averaging the predictability times over several 
realizations, see equation (\ref{fsle:map}). In general, one must choose 
$\delta_{min} \ll \delta_{0}$, in order to allow the direction of the 
initial perturbation to align with the most unstable direction in the 
phase-space. Moreover, one must pay attention to keep 
$\delta_{M} < \delta_{saturation}$, so that all the thresholds 
can be attained. 

Our hope is that, as in the usual hydrodynamical description of 
thermodynamic systems, the variables of systems with macroscopic 
chaos can be separated in slow variables (given by $m(t)$, $\sigma(t)$
and so on) and fast variables.

We have computed the FSLE (\ref{fsle}) looking only at $m(t)$ 
both for the homogeneous GCM (\ref{eq:uno}), 
using as local map the tent map, and for the heterogeneous 
GCM (\ref{eq:dueuno}), for different values of the parameters.
Practically the numerical experiments has been performed as follows:
we have averaged, over many realizations, the predictability 
times $T_{r}(\delta_{n})$ (defined above) related to $|\delta m(t)|$  
which is initialized at the value $\delta m(t)=\delta_{min}$.
This has been done considering, for each realization, the perturbed 
system that is obtained by shifting all the elements of the unperturbed 
one by the quantity $\delta_{min}$ (i.e. $x^{'}_{i}(0)
=x_{i}(0)+\delta_{min}$). 

As can be seen from figure 5a, which shows $\lambda(\delta)$ versus
$\delta$ in the case with macroscopic chaos, two plateaux exist: at
small values of $\delta$ ($\delta \leq \delta_{1}$) one has, as
expected from general considerations, $\lambda(\delta)=\lambda_{1}$;
for $\delta \geq \delta_{2}$ one has another plateau which can fairly
be called ``macroscopic'' Lyapunov exponent,
$\lambda(\delta)=\lambda_{macro}$.  Moreover, $\delta_1$ and
$\delta_2$ decrease at increasing $N$. In fact, from figure 5b, where 
we report $\lambda(\delta)$ versus $\delta \sqrt{N}$
for different values of $N$, one can see that both $\delta_{1}$ and
$\delta_{2}$ scale as $\sim 1/\sqrt{N}$: by rescaling $\delta$
with $\sqrt{N}$, one observes that $\delta_1 \sqrt{N}$
(which indicates the end of the microscopic plateau) and $\delta_2
\sqrt{N}$ (which marks the end of the saturation due to the
microscopic dynamics and the beginning of the macroscopic plateau) 
coincide for all values of $N$. Incidentally, one can also observe that the
macroscopic plateau, being almost non-existent for $N=10^4$, becomes
more and more resolved and extended on large values of $\delta
\sqrt{N}$ at increasing $N$ up to $N=10^7$. From these features we can
argue that the macroscopic motion is well defined in the limit $N
\rightarrow \infty$ and therefore we can conjecture that in this limit
the microscopic signature in the evolution of $\delta m(t)$ completely
disappears in favor of the macroscopic behaviour.

It is important to note that in order to obtain the results shown in
fig.5 and fig.7 some precautions must be taken in initializing the
system.  In fact, because of the complex structure of the phase space of
this system, one has that different initial conditions can led to
different attractors \cite{Kan95}. Therefore, choosing random initial 
conditions while increasing $N$, one likely obtains different values
for the macroscopic and/or microscopic Lyapunov exponents. In order to
maintain the system on the same attractor, we kept fixed the
percentage of elements {\it per} band (see Kaneko \cite{Kan95}),
at varying the number of elements $N$.

In the case of standard chaos ($\lambda_{macro}<0$)
one has only the microscopic plateau and then a fast decreasing of
$\lambda(\delta)$, as can be seen in fig. 6. Also in the cases of non
trivial macroscopic behaviour, either periodic or quasi-periodic
(i.e. $\lambda_{macro}=0$), we see only the microscopic plateau
followed by a decreasing up to the saturation.  This is a consequence
of the fact that $\lambda(\delta)$ is not able to distinguish between
$\lambda <0$ and $\lambda=0$, by definition, see eq. (\ref{fsle}).

One can wonder if it is possible to compute the macroscopic Lyapunov
exponent directly looking at $\langle \ln |\delta m| \rangle$
versus $t$. The macroscopic regime is expected to appear as a change
of slope when the size of the perturbation is $O(1/\sqrt{N})$ (as
discussed in sect. \ref{sec:3} in the case of standard chaos).  In
fig. 7 $\langle \ln |\delta m| \rangle$ is shown as a function of $t$
for the same systems of fig. 5 (with identical initial
conditions). One can see that for small values of $\delta m(t)$ one
has a well defined exponential growth according to the microscopic
Lyapunov exponent. This regime stops at $\delta m(t) \sim
O(1/\sqrt{N})$ i.e.  when the microscopic error growth saturates. At
larger values of $\delta m(t)$ one observes a second (macroscopic)
slope. One has a clear evidence also from fig. 7 that the macroscopic
motion sets in at a scale that is smaller when $N$ is larger, as already
established by the analysis of fig. 5. In the case of fig. 7 one has
also a fair quantitative agreement with the results obtained with the
FSLE method (fig. 5). However in general, because of the
intermittency, the macroscopic regime can start at different times and
with different rates, therefore the naive method in which the averages
are performed at fixed delay times can give spurious laws; see ref
\cite{ABCPV96,ABCCV97} for a discussion of these effects.
 
Of course {\it a priori} there is no deep reason to believe that 
the above analysis performed only on one variable, $m(t)$, is enough 
accurate, therefore we performed a similar study looking at two 
macroscopic variables, $m(t)$ and $\sigma(t)$. Since there is not 
a natural way to introduce a perturbation on $m(t)$ and
$\sigma(t)$, we are forced to follow the method of Wolf et al. \cite{Wolf85} 
for the computation of the Lyapunov exponent from data, but with 
some modifications. We considered the phase space generated by 
the variables $m(t)$, 
$\sigma(t)$ and we looked at the time separation between 
{\it analogues} whose initial
distance $\delta_{0}$ is not very small (i.e. in the region of the 
``macroscopic'' plateau in fig. 5). Following the evolution of the 
full system we compute the quantity:
\begin{equation}
\Gamma(\tau)=\frac{1}{\cal A} \sum_{j=1}^{\cal A} 
\ln \frac{\Delta(t_{j}+\tau)}{\Delta(t_{j})}\,,
\label{eq:dav1}
\end{equation}
where $\Delta(t)=\sqrt{\delta m^{2}(t)+\delta \sigma^{2}(t)}$, 
$t_{j}$ are the ``return'' times such that 
$\Delta(t_{j}) \in [\delta_{0},2\,\delta_{0}]$ and ${\cal A}$ is 
the number of analogues. Fig. 8 shows $\Gamma(\tau)$ versus $\tau$.

It is important to stress the relevance of the value of $\delta_0$,
which must be not too small. Using a $\delta_0$ out of the 
``macroscopic'' (i.e. $\delta_{0} < \delta_{2}$) plateau one has spurious 
behaviour due to intermittency (see ref. \cite{ABCCV97}).

It is remarkable the agreement between the values for the macroscopic
Lyapunov exponent obtained with the two methods.

\section{Conclusion and discussion}
\label{sec:5}

In this paper we proposed a quantitative characterization of the 
coherent evolution in globally coupled maps that show {\it macroscopic 
chaos}. From a conceptual point of view, the irregular behaviours of 
global observables -- which typically evolve on time scales much longer 
than those related to the evolution of a single element $x_{n}(t)$ --
are for the GCM the analogous of the chaos for hydrodynamical 
equations in the description of the molecular motion.

In generic systems there are not a systematic method to obtain the 
equations that describe the evolution of the degrees of freedom 
accounting for the large scale behaviour. Therefore a straightforward 
application of the usual dynamical systems methods is not able to 
give suitable characterizations. In particular, the usual Lyapunov 
exponent, which gives a microscopic description, is not able to 
capture the essential features of this macroscopic phenomenon.

It is worth to note that in the cases where it is possible to derive
hydrodynamical-like equations, such as the Perron-Frobenius equation,
one obtains a description of the system at a macroscopic
level, see for a detailed discussion \cite{Kan95,PK94}. We have shown
that, in the cases in which this approach is missing, the recently
introduced Finite Size Lyapunov Exponent \cite{ABCPV96} is able to
characterize, in a consistent way, the macroscopic behaviour of the
system and gives a measure of the chaotic motion at a coarse-grained
level. We have the following scale-dependent scenario for the
mechanism acting on the evolution of the perturbation of global
observables (i.e. the mean field):
\begin{itemize}
\item 
at small values of the perturbation ($\ll 1/\sqrt{N}$) 
we recover the microscopic Lyapunov exponent, simply because at 
these scales the evolution of the perturbation is driven by 
the microscopic dynamics; 
\item 
at larger values ($\gg 1/\sqrt{N}$) we obtain a macroscopic 
Lyapunov exponent, which can be much smaller than the first 
Lyapunov exponent of the microscopic dynamics.
\end{itemize}

The above scenario is confirmed using (with some important 
changes) the method of Wolf et al. \cite{Wolf85} for the 
computation of the Lyapunov exponent from  numerical data. 
Basically, in presence of macroscopic chaos one observes that: 
the variables of the system are separated in slow variables 
(e.g. $m(t), \sigma (t)$) and fast variables (see fig.s 5 and 7). 
For a discussion of the FSLE in systems with slow and 
fast dynamics see Ref. \cite{BGPV98}.  

Let us stress that also with a naive numerical treatment of the
one-body description one can observe a rather clear evidence for the
two regimes (the microscopic and the macroscopic ones) characterized
by two very different ``Lyapunov exponent''.  But, as discussed in the
Appendix, the problem of the choice of the norm (that is present also
working with the histograms) gives some unavoidable troubles which do
not allow for a detailed comparison with the other methods.

Finally, it is interesting to mention the case of ``single-band''
chaos (see ref. \cite{Kan95}), where one has irregular behaviour with
some macroscopic coherence, but there is not a separation between the
microscopic and the macroscopic scales. The microscopic Lyapunov
exponent has the same value of the macroscopic one
($\lambda_{macro}=\lambda_{1}$). This means that in this case 
-- as  fig. 9 shows -- $\lambda(\delta)$ has a single plateau, 
extending over a range of scales larger than that of the 
standard chaos (fig. 6), so as to include the macroscopic scales.
Let us remark that this case of ``single-band''
chaos is reminiscent of the two-dimensional turbulence \cite{ABCPV97},
where the $\lambda(\delta)$ does not depend on $\delta$ in a very
large range of values; that is, a broad range of scales have equal
characteristic times.

\hfill\break

After the submission of this paper, the authors are informed of the 
recent preprint by T. Shibata and K. Kaneko ``Collective Chaos'' 
(chao-dyn/9805009), where a related study is presented. In particular, 
T. Shibata and K. Kaneko use the FSLE to study the macroscopic 
properties of a heterogeneous GCM. Their results are in agreement 
with ours. 

\acknowledgments 

We thank E. Aurell and L. Biferale for useful suggestions and a
careful reading of the paper. This paper has been partially supported
by INFM (PRA-TURBO) and MURST (program 9702265437). 

\appendix
\section*{}
 
One could naively think that the behaviour of $|\delta m(t)|$
and $||\delta \rho_{t}||$ give the same information.
In this appendix we show that the above idea does not work because of 
the unavoidable troubles related to the choice of a norm for 
$\delta \rho_{t}$.

In order to show this we discuss two explicit cases which clarify 
this point. As first example we consider $\rho(x)$ uniform in the 
interval $[a,b]$ and the perturbed distribution $\rho^{'}(x)$ uniform
in the interval $[a+\delta m,b+\delta m]$, i.e. with a shifted average. 
Let us now compute two different norms of $\delta \rho= \rho^{'}-\rho$:
\begin{equation}
||\delta \rho||_{1}=\int |\delta \rho(x)| {\rm d}x= \frac{2}{b-a} |\delta m|
\label{eq:a1}
\end{equation}
\begin{equation}
||\delta \rho||_{2}=\sqrt{\int \delta \rho^{2}(x) {\rm d}x}= 
\sqrt{\frac{2}{(b-a)^{2}} |\delta m|}\,.
\label{eq:a2}
\end{equation}
Assuming that $\delta m$ is related to a GCM system, it is clear that
the computation of the Lyapunov exponents according to the norms 
defined above gives, in this case (uniform distribution),
two Lyapunov exponent different by  a factor two.
Nevertheless with different shape of the distribution we have 
different results. To show this last issue we consider now a 
$\rho(x)$ with a symmetric triangular shape in the interval $[a,b]$
and zero for $x<a$ or $x>b$. The perturbed $\rho^{'}$ has the same 
functional form with the average shifted of $\delta m$.
The same computations as before lead to:
\begin{equation}
||\delta \rho||_{1} =(b-a) |\delta m|+O(|\delta m|^{2})
\label{eq:a3}
\end{equation}
\begin{equation}
||\delta \rho||_{2}=\sqrt{\frac{b-a}{2}}|\delta m|+O(|\delta m|^{2})\,.
\label{eq:a4}
\end{equation}

 From the above examples it is clear that the relation between $|\delta m|$
and $||\delta \rho||$  depends dramatically on the explicit norm used 
and on the shape of the $\rho$. In addition, in the real time 
evolution of $\delta \rho_{t}$ one has not only shifting of the average 
but also variations of $\sigma^{2}$ and other parameters of the 
distribution that complicate the interpretation of $||\delta \rho||$
in terms of $|\delta m|$. Therefore there is no reason to expect
a quantitative agreement between numerical study (using histograms) of the 
one-body description and $|\delta m(t)|$.

\newpage

\centerline{\bf FIGURE CAPTIONS}

\begin{description}

\item{FIGURE 1:}  (a) Plot of $m(t)$ versus $m(t-1)$ for the system 
(\ref{eq:uno}) with $a=1.7$ $\epsilon=0.3$ and $N=10^{6}$; fig (1b)
is an enlargement of (a).

\item{FIGURE 2:}  Probability distribution function $P_{n}(x)$ for 
two different $x_n$ for the system eq. (\ref{eq:uno})
with $a=1.8$, $\epsilon=0.3$ and $N=10^5$. The distribution are 
computed following $x_{n}(t)$ for $5\cdot 10^5$ temporal steps.

\item{FIGURE 3:} Absolute value of the correlation function
$C_{macro}(\tau)$ (solid line) compared with the absolute value of the
correlation functions for two different $x_{i}(t)$,
$C^{i}_{micro}(\tau)$, (dashed and dot-dashed lines). The parameters are
chosen as in fig. 2.

\item{FIGURE 4:}  $<\ln |\delta m| >$ versus $t$ for 
the system (\ref{eq:uno}) with $a=1.8$, $\epsilon=0.08$ and $N=10^{4}$.
The squares are for an initial perturbation 
$\delta m(0)=10^{-6} \ll 1/\sqrt{N}$,  the circles for 
$\delta m(0)=0.15 \gg 1/\sqrt{N}$, and the dotted line 
has slope $\lambda_{micro}$. The average is over $10^3$ realizations.

\item{FIGURE 5a:} $\lambda(\delta)$ versus $\delta$ for the system
(\ref{eq:uno}) with $a=1.7$, $\epsilon=0.3$ for $N=10^4$ ($\times$),
$N=10^5$ ($\Box$), $N=10^6$ ($\odot$) and $N=10^{7}$ ($\triangle$) .
The first plateau corresponds to the microscopic Lyapunov exponent
$\lambda_{micro} \approx 0.17$ and the second one to the macroscopic
Lyapunov exponent $\lambda_{macro}\approx 0.007$.  The average is over
$2\cdot 10^3$ realizations for $N=10^{4},\,10^5,\,10^6$ and $250$
realizations for $N=10^7$. The system has been initialized in order
to leave the attractor unchanged (as explained in the text).

\item{FIGURE 5b:} The same as figure 5a rescaling the $\delta-$axis
with $\sqrt{N}$.

\item{FIGURE 6:}  $\lambda(\delta)$ versus $\delta$ for 
the system (\ref{eq:uno})
with $a=1.8$, $\epsilon=0.08$ for $N=10^4$ ($\Box$), $N=10^5$ ($\odot$) and
$N=10^{6}$ ($\triangle$). The plateau corresponds to the microscopic
Lyapunov exponent $\lambda_{micro} \approx 0.5$.
The average is over $2\cdot 10^3$ realizations.

\item{FIGURE 7:}  $<\ln |\delta m| >$ versus $t$ for 
the system (\ref{eq:uno}) with $a=1.7$, $\epsilon=0.3$ and $N=10^{4}$
(upper curve), $N=10^{5}$ (middle curve), $N=10^{6}$ (lower curve) 
The straight lines correspond to the two exponential 
growth the microscopic one ($\lambda_{micro} \approx 0.17$) 
and the macroscopic ($\lambda_{macro}\approx 0.01$).
The average is over $500$ realizations.

\item{FIGURE 8:} $\Gamma(\tau)$ versus $\tau$ for the system
(\ref{eq:uno}) with $a=1.7$, $\epsilon=0.3$ and ${\cal A}=250$, the
dashed line corresponds to $\lambda_{macro}\approx 0.006$. The
distance has been plotted every two time steps to eliminate
oscillations given by the swiching between bands.

\item{FIGURE 9:}  $\lambda(\delta)$ versus $\delta$ for 
the system (\ref{eq:uno})
with $a=1.5$, $\epsilon=0.3$ for $N=10^4$ ($\Box$), $N=10^5$ ($\odot$). 
The plateau corresponds to a Lyapunov exponent 
$\lambda_{micro} \approx 0.28$.
The average is over $2\cdot 10^3$ realizations.

\end{description}
%\end{document}

%%%%%%%%%%%%%%%%%%%%%%%%%%%%%%%%%%%%%%%%%%%%%%%%%%%%%%%%%%%%%%%%%
%%%%%%%%%%%%%%% 	FIGURE		%%%%%%%%%%%%%%%%%%%%%%%%%
%%%%%%%%%%%%%%%%%%%%%%%%%%%%%%%%%%%%%%%%%%%%%%%%%%%%%%%%%%%%%%%%%

\newpage

\begin{figure}[hbt]
\epsfbox{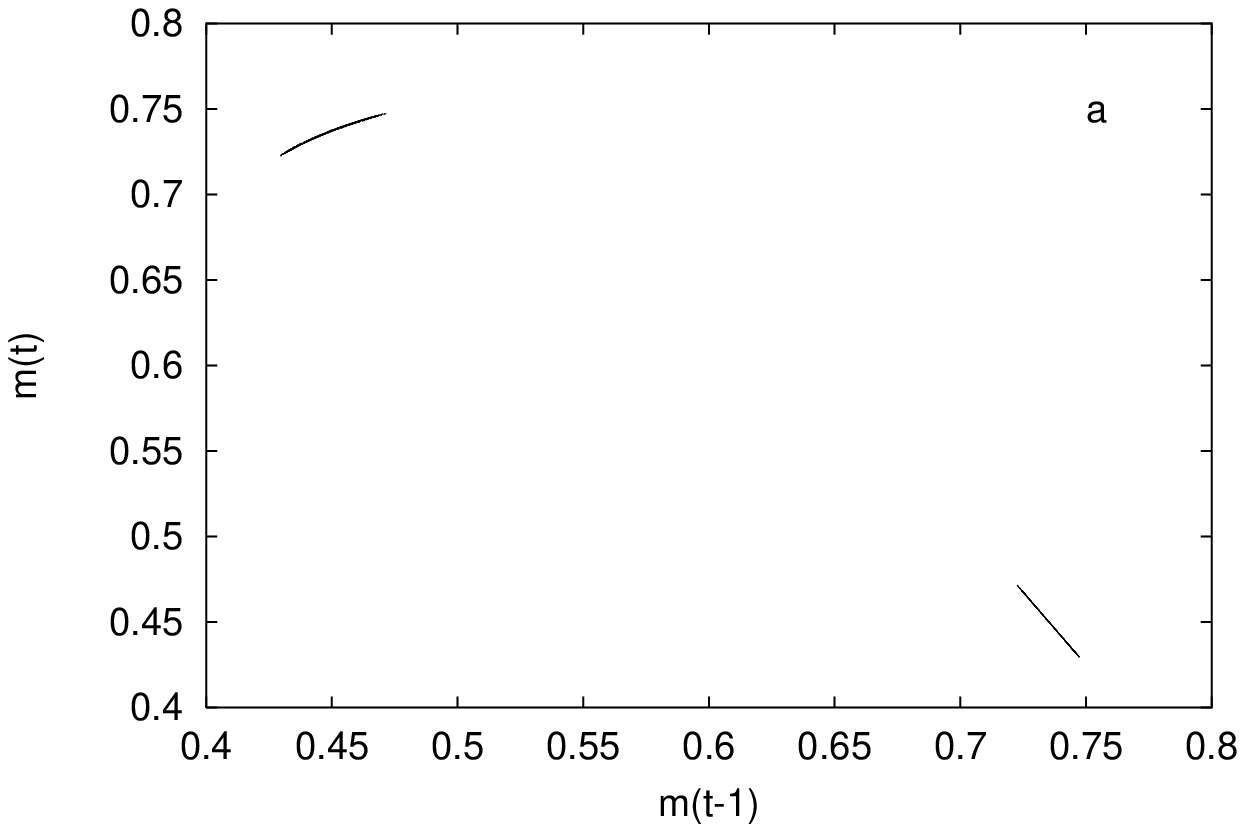}

\epsfbox{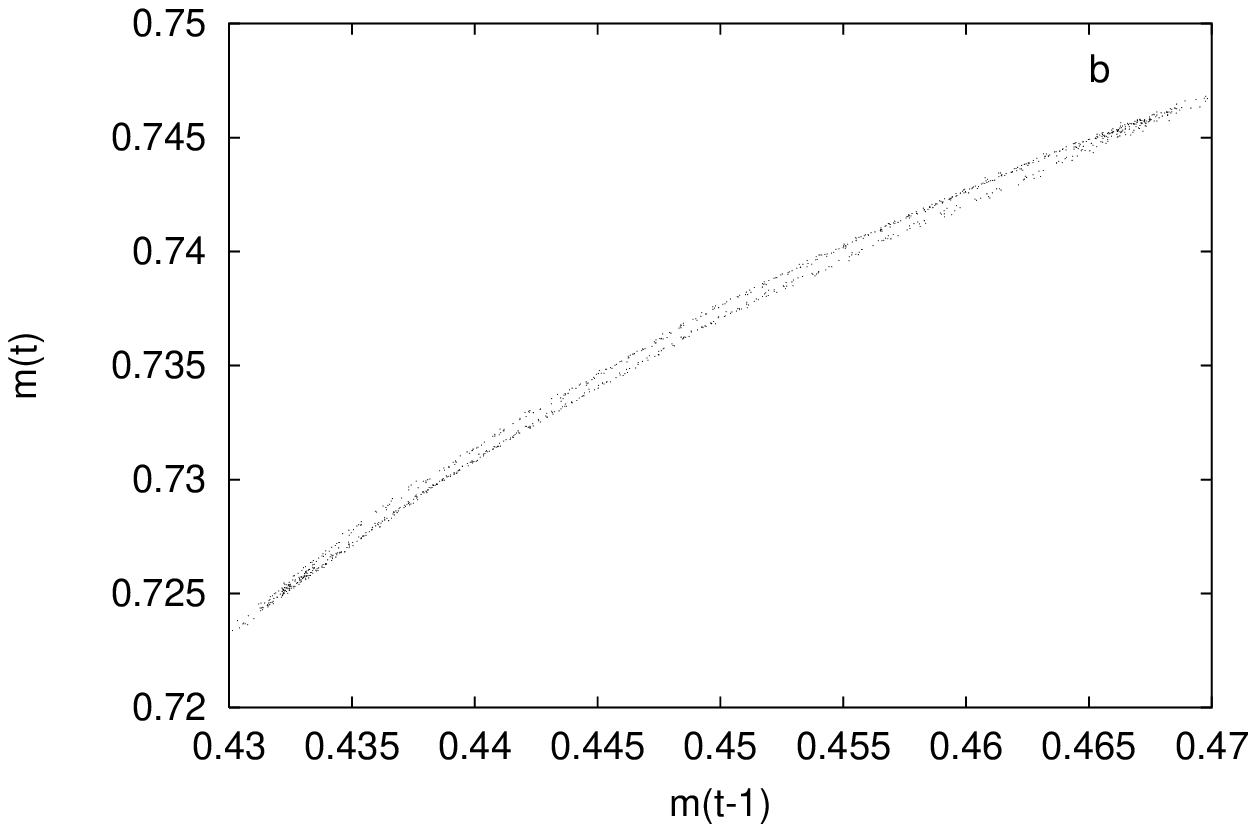}
\caption{}
\label{fig1}
\end{figure}

\newpage

\begin{figure}[hbt]
\epsfbox{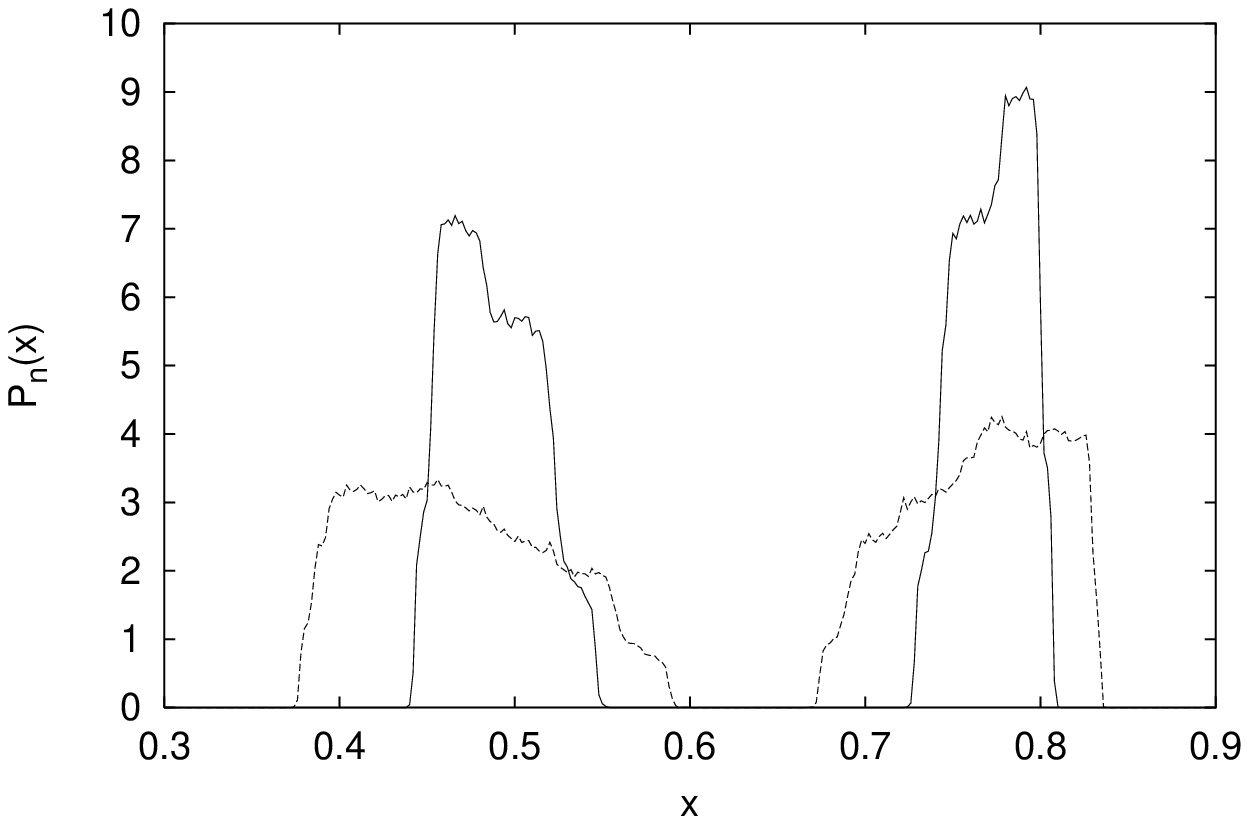}
\caption{}
\label{fig2}
\end{figure}

\newpage

\begin{figure}[hbt]
\epsfbox{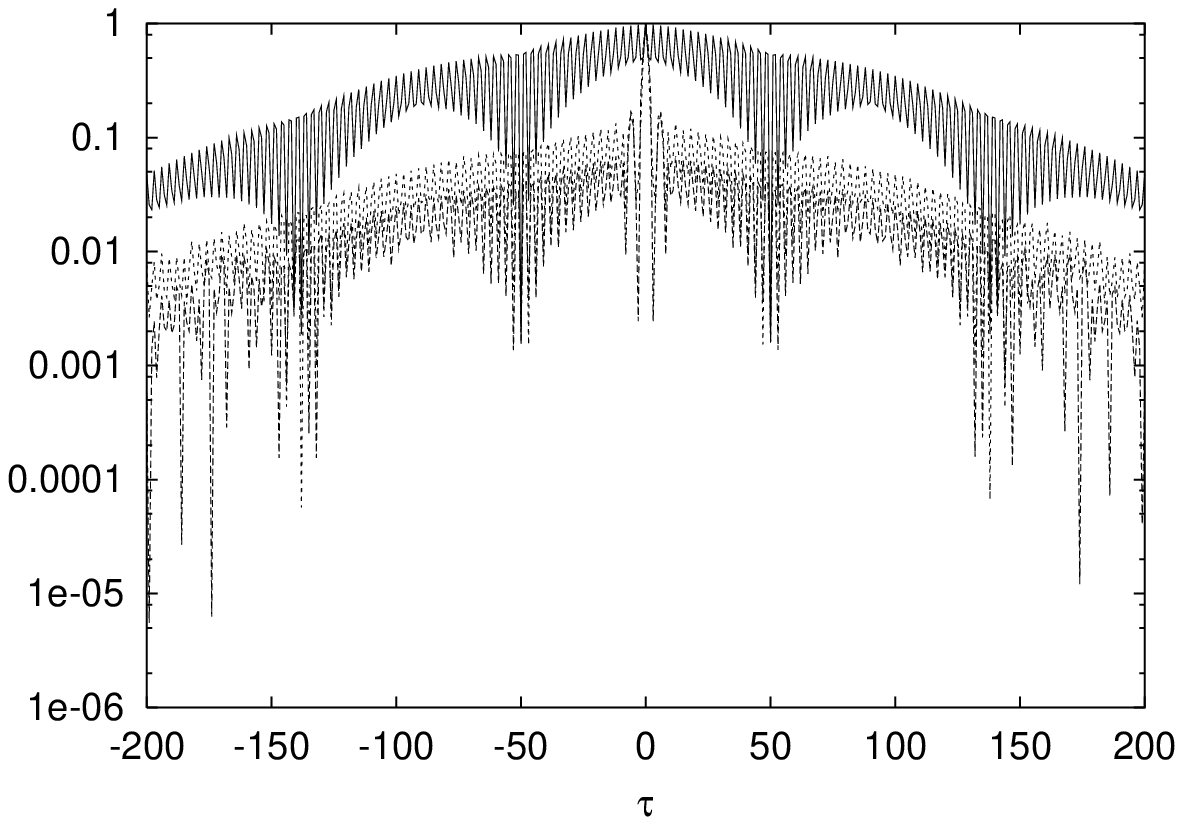}
\caption{}
\label{fig3}
\end{figure}

\newpage

\begin{figure}[hbt]
\epsfbox{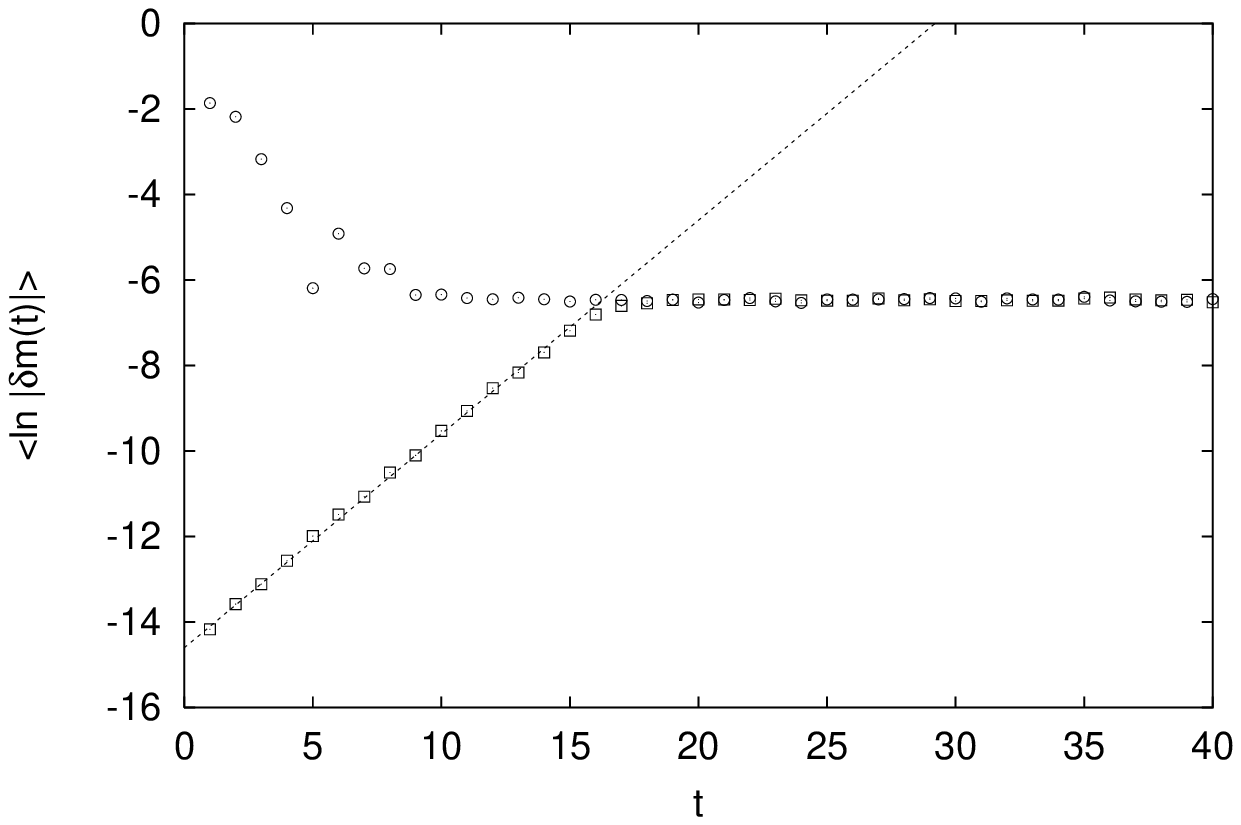}
\caption{}
\label{fig4}
\end{figure}

\newpage

\begin{figure}[hbt]
\epsfbox{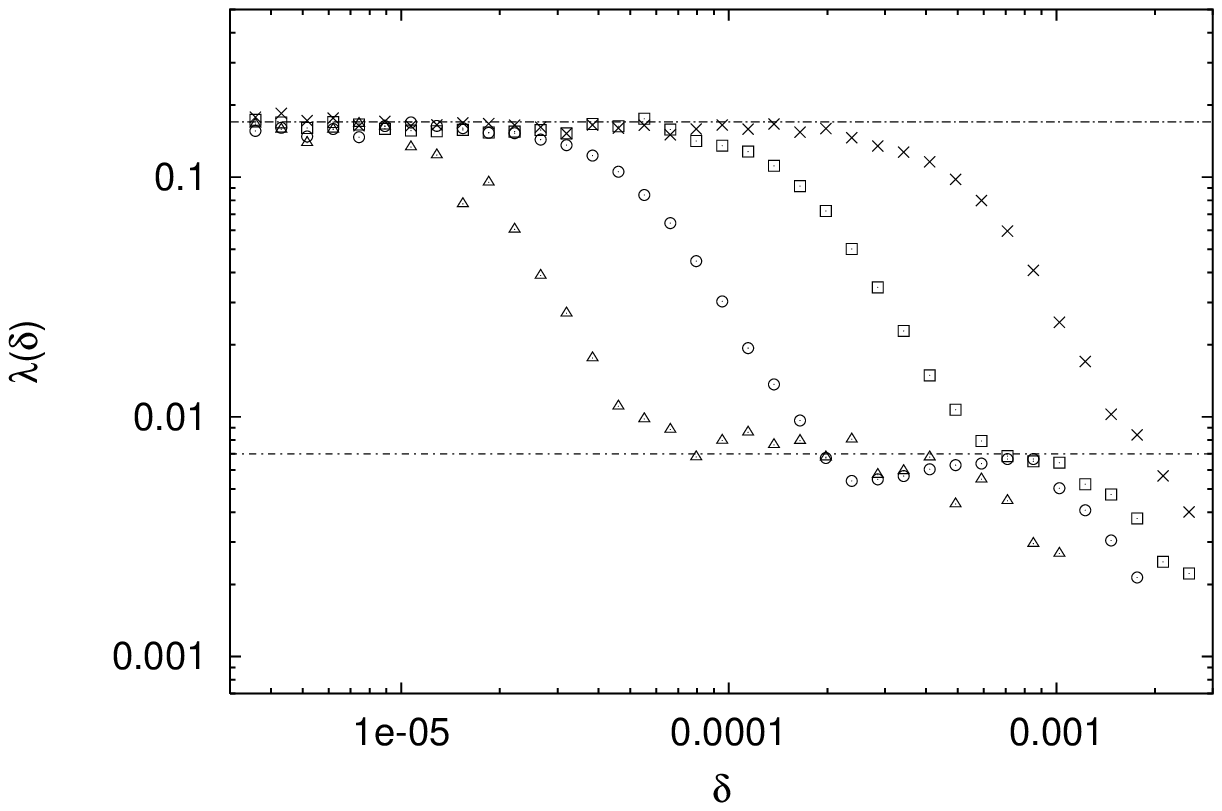}

\epsfbox{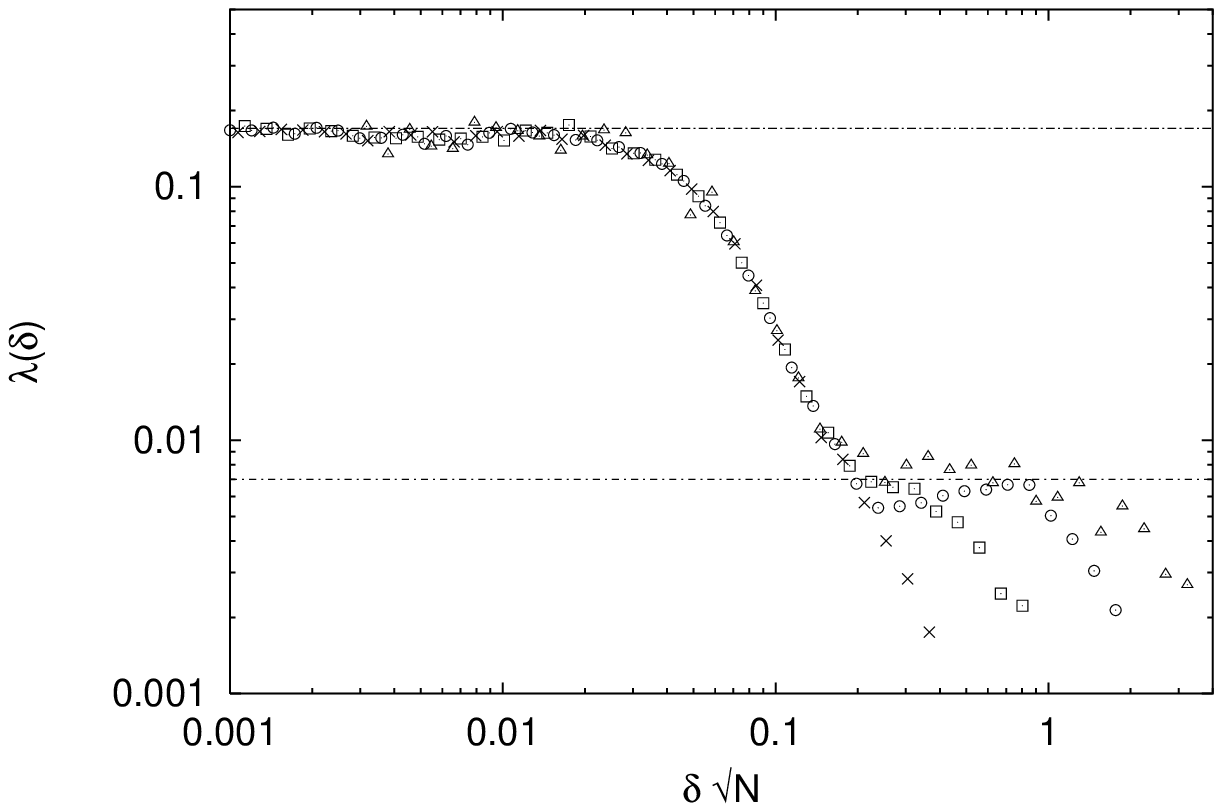}
\caption{}
\label{fig5}
\end{figure}

\newpage

\begin{figure}[hbt]
\epsfbox{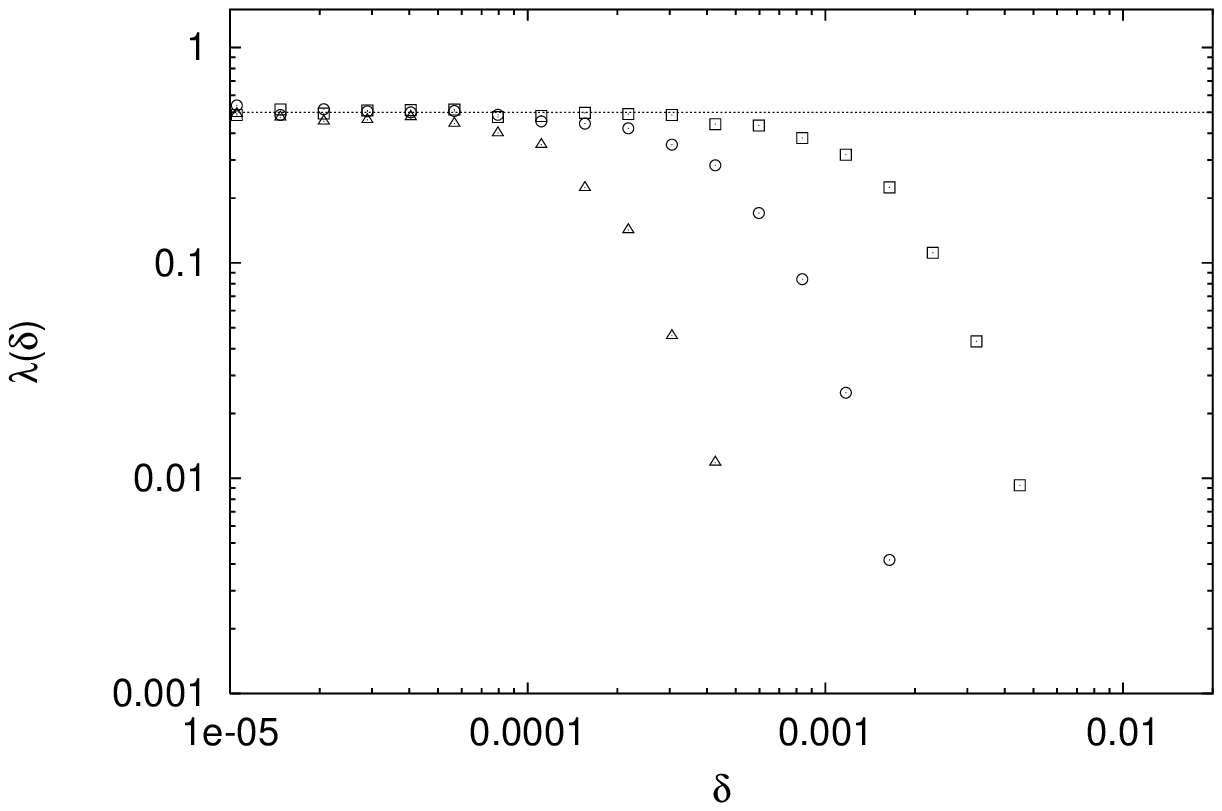}
\caption{}
\label{fig6}
\end{figure}

\newpage

\begin{figure}[hbt]
\epsfbox{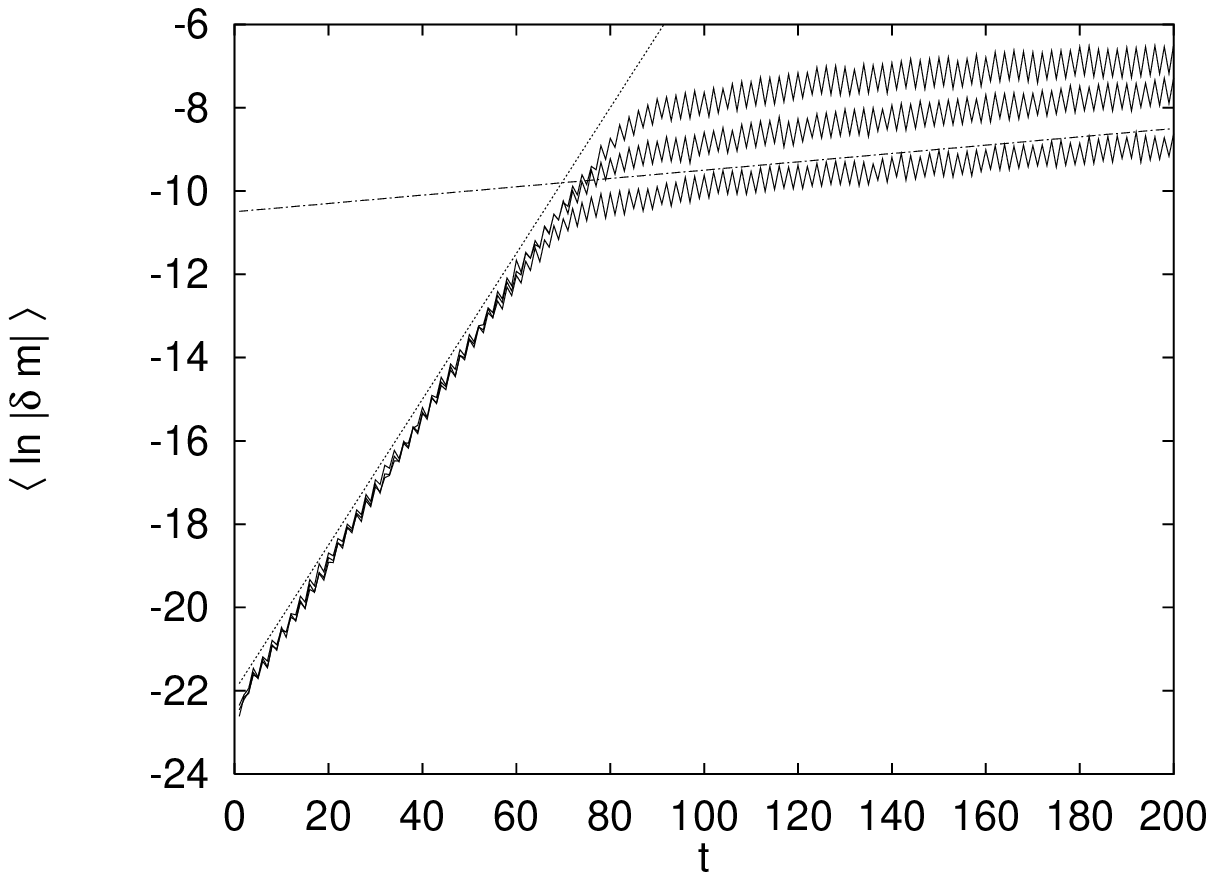}
\caption{}
\label{fig7}
\end{figure}

\newpage

\begin{figure}[hbt]
\epsfbox{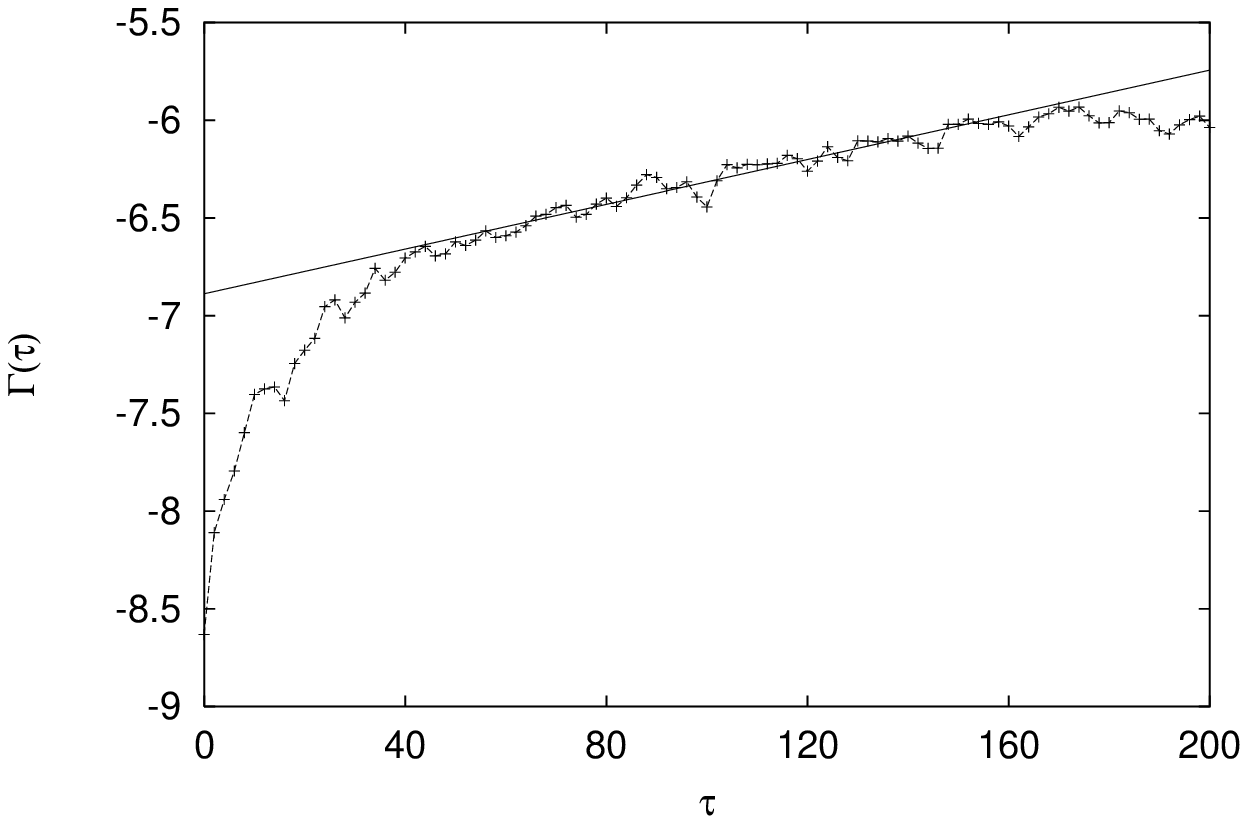}
\caption{}
\label{fig8}
\end{figure}

\newpage

\begin{figure}[hbt]
\epsfbox{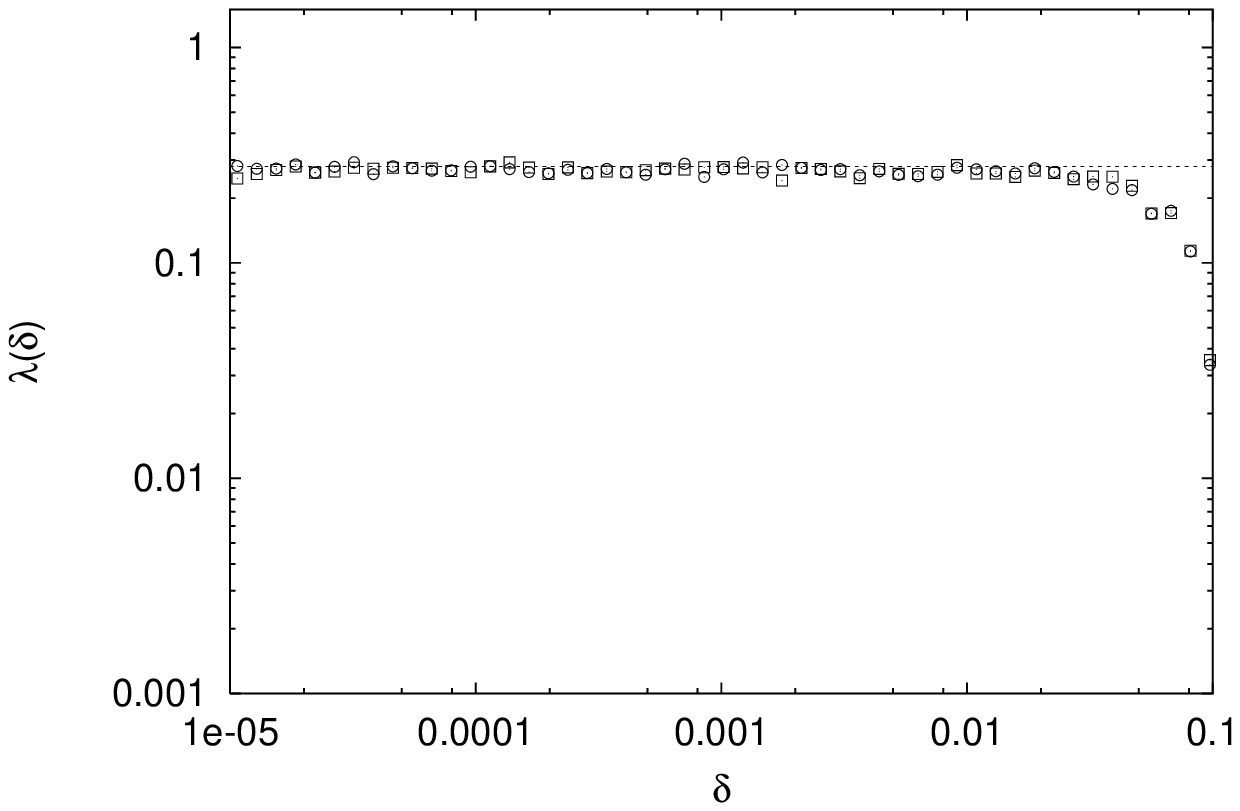}
\caption{}
\label{fig9}
\end{figure}

\end{document}